\documentstyle[11pt,aspconf]{article}
\def\lax    {${_<\atop^{\sim}}$}
\def\gax    {${_>\atop^{\sim}}$}

\begin{document}

\title{The Evolving XUV Absorber in NGC3516}
\author{Smita Mathur, Belinda Wilkes \& Thomas Aldcroft}
\affil{Harvard-Smithsonian Center for Astrophysics, 60 Garden St.,
Cambridge, MA 02138, USA}

\begin{abstract}
For NGC3516 we find that the X-ray warm absorption and the broad UV
associated absorption features can be produced by the same absorbing
material. We argue that the evolution of the XUV absorber from
pre-1992 to 1995 is consistent with expectation for an expanding,
outflowing material.
\end{abstract}


\section{NGC3516: X-ray and UV Observations}

 NGC3516 contains the strongest UV absorption system known in a
Seyfert 1 galaxy. This system contains at least two distinct
components: a broad (FWHM$\sim$2000 km/s) variable component and a
narrow ($\sim$500 km/s) non-variable component.  Both the broad and
the narrow systems contain high as well as low ionization absorption
lines. Recent observations have shown that the broad high ionization
absorption lines have {\bf \it disappeared} since $\sim$1992 (Koratkar
et al.  1996 and references therein, Kriss et al. 1996)

 We analyze a high signal-to-noise (S/N) ROSAT PSPC archival spectrum
of NGC3516 obtained in 1992.  The high S/N allows the strong detection
of both OVII and OVIII edges independently, in spite of the limited
spectral resolution of the PSPC. A warm absorber fit to the data shows
that the absorber is highly ionized (U$=10^{+2.6}_{-2.1}$), and has a
large column density N$_H \sim 10^{22}$ cm$^{-2}$.

\section{The XUV Absorber}

 In several AGN, the X-ray and the UV absorbers were found to be one
and the same (the `XUV' absorbers, Mathur et al. 1994, 1995). Is it
also true for NGC3516? The absorption systems in NGC3516 are clearly
complex with multiple components (Kriss et al. ~1996). It is the {\it
high ionization, broad} absorption system that is most likely to be
associated with the X-ray warm absorber.  Investigation of this
question is tricky, however, because the broad absorption lines have
disappeared. Here we argue that the XUV absorption picture is {\it
consistent} with the presence of a highly ionized X-ray absorber and the
current non-detection of CIV and NV broad absorption lines (see Fig. 1).
The X-ray absorber {\it MUST} have a UV signature showing OVI
absorption lines (Fig. 1). Since there were no simultaneous
ROSAT \& far-UV observations in 1992, this cannot be directly determined.
However, we note that in the 1995 HUT observations, OVI doublets are
unresolved, but consistent with being  broad
(FWHM=1076$\pm$146 km/s, Kriss et al. 1996).
\begin{figure}
\vspace*{-0.9in}
\centerline{
\epsscale{.75}
\plotone{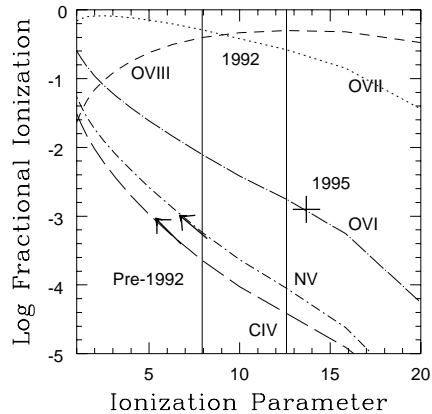}
}
\vspace*{-0.7in}
\caption{Ionization fractions f of OVI, OVII, OVIII, CIV and NV as a
function of ionization parameter, U (with CLOUDY, Ferland 1991). The
vertical lines define the range of U for which the ratio
f$_{\mbox{OVII}}$/f$_{\mbox{OVIII}}$ lies within the observed ROSAT
range. The arrows on the CIV and NV curves indicate the lower limits
of f$_{\mbox{CIV}}$\gax $3\times 10^{-4}$ and f$_{\mbox{NV}}$\gax
$3.1\times 10^{-4}$ based on the published IUE data.  The + mark
corresponds to the HUT data in Kriss et al. 1996.}
\end{figure}

 We argue that the XUV absorber in NGC3516 has evolved with time (Fig.
1). {\bf Pre-1992:} It showed broad, high ionization CIV and NV
absorption lines and an X-ray ionized absorber (U \lax 7). As it
evolves, outflowing and expanding, the density falls and the
ionization parameter increases. {\bf 1992:} CIV and NV absorption lines
disappeared; X-ray absorber is still present with OVI lines in the UV
(U$\sim$10) (No UV data available to verify). {\bf 1995:} CIV and NV
absorption line remain absent; X-ray absorber is present. OVI lines
are present, and detected with HUT (U$\sim$13.5). {\bf Post-1996:} We
predict that the OVI absorption lines will disappear as the ionization
parameter increases further (U\gax 20). The OVIII edge will continue
to strengthen relative to the OVII edge.  Eventually, even the X-ray
absorber will also disappear.

\acknowledgments
SM gratefully acknowledges the financial support of NASA grant
NAGW-4490 (LTSA) and BW, TA of NASA contract NAS8-39073 (ASC).

\end{document}